\documentclass[twocolumn]{aastex62}
\usepackage{amsmath}
\graphicspath{{./}{figures/}}

\received{June 19, 2019}
\revised{June 26, 2019}
\accepted{July 01, 2019}
\shorttitle{Habitability of Teegarden’s Star planets}
\shortauthors{Wandel \& Tal-Or}
\begin{document}

\title{On the Habitability of Teegarden's Star planets}

\correspondingauthor{Amri Wandel}
\email{amri@mail.huji.ac.il}

\author{Amri Wandel}
\affil{Racah Institute of Physics, Faculty of Natural Sciences, The Hebrew University of Jerusalem, Jerusalem, Israel}

\author{Lev Tal-Or}
\affiliation{Department of Geophysics, Raymond and Beverly Sackler Faculty of Exact Sciences, Tel Aviv University, Tel Aviv, Israel}



\begin{abstract}

We study the habitability of the two $1.3_{-0.3}^{+0.7}$ Earth-mass planets, recently detected by the CARMENES collaboration, around the ultra-cool nearby M-dwarf Teegarden's Star. With orbital periods of $4.9$ and $11.4$ days, both planets are likely to be within the habitable zone and tidally locked. They are among the most Earth-like exoplanets yet discovered. Applying an analytic habitability model we find that surface liquid water could be present on both planets for a wide range of atmospheric properties, which makes them attractive targets for biosignature searches. The prospects of the planets retaining such an atmosphere over their history are discussed.

\end{abstract}




\section{Introduction} \label{sec:intro}

The CARMENES collaboration has recently announced (Zechmeister et al. 2019, hereafter Z19) the detection of two planet candidates, each with $1.0$--$2.0$\,M$_{\oplus}$ mass, orbiting the ultra-cool (M7V) nearby ($3.83$\,pc) Teegarden's Star (hereafter TG). Assuming the estimated host's mass of $0.089\pm 0.009$\,M$_{\odot}$, their periods, $4.91$\, and $11.4$\,days, correspond to an orbital distance of $0.025$ and $0.044$ AU, $\pm3$\,$\%$, respectively. At these distances, with the estimated host's age of $>8$\,Gyr (Z19), both planets are most probably tidally locked (e.g., Griessmeier et al. 2009).

We investigate the habitability of the exoplanets Teegarden's Star b and c (hereafter TGb and TGc, respectively), and their potential to have an atmosphere that would support surface liquid water, using an atmospheric habitability model for locked planets (Wandel 2018, hereafter W18). We derive the dependence of the surface temperature of the planets on circulation and greenhouse heating, as well as their habitability regime in the atmospheric parameter space.

%%===============================================
\section{Habitability ranges for the planets of Teegarden's Star} \label{sec:2}

\subsection{The analytic climate-habitability model}

Planets within the habitable zone (HZ) of late M-dwarfs are tidally locked (Griessmeier et al. 2009). Without atmosphere the surface temperature would be determined only by the irradiation from the host star and the  angular distance from the substellar point. W18 puts together an analytic 1D model, in which the atmosphere's impact on the surface temperature distribution is evaluated, taking into account (a) the irradiation from the host star (insolation), (b) atmospheric transmission (screening and greenhouse effect), and (c) horizontal heat transport due to circulation, convection, and advection. The values (b) and (c) can in principle be calculated, given the planet's data (specific gravity, rotation) and the atmospheric properties (composition, pressure, heat capacity,  wind speed, global circulation patterns, etc.). However, as these data are difficult to obtain and disentangle in exoplanets, they are parameterized using the atmospheric heating and the global redistribution factors, respectively.
The model combines elements from previous analytic temperature models of locked planets (e.g., Haberle et al. 1996; Koll and Abbot 2016), but the model's novelty and strength is in being independent of the specific atmospheric composition and of the details of the energy transport mechanism, both being represented in a parametric form: the former by the atmospheric heating factor and the latter by the global heat redistribution parameter. This, of course, is also a weakness, when it comes to subjects like 2D and 3D effects, flow patterns like cells and vertical transport, and more complex feedback mechanisms that depend on the composition, like clouds (e.g., Yang et al. 2013). 
The analytic expression of the surface temperature is combined with the temperature boundaries of the HZ, to define a habitability range in the two-dimensional parameter plane, namely, atmospheric heating and circulation (e.g., Fig. 2). This will be discussed in more detail in subsection 2.3.

\subsection{Surface temperature distribution}

Following W18 we define the dimensionless heating factor $H$ which is a measure of the surface heating, combining the host's irradiation with the albedo ($A$), the atmospheric screening ($\alpha$), and the greenhouse heating factor ($H_g$),
\begin{equation}
H = (1-A)\,H_g\,\alpha\,S / S_{\oplus} = H_{atm}\,s,
\end{equation}
where $s = S / S_{\oplus}$ is the insolation relative to Earth. The product $(1-A)H_g\alpha = H_{atm}$ is defined as the atmospheric heating factor. $H_g$ is also related to the atmospheric optical depth in the lower wavelength band (IR) by $H_g \approx \tau _{\rm LW} + 1$. Typical values of the heating parameter for the solar system are: $H\sim1$ (Earth), $\sim0.3$ (Mars), and $\sim50$ (Venus). For a locked planet, the surface temperature distribution can be calculated for each "latitude" (angular distance from the substellar point) by equating the local heating and cooling. In the model this is combined with the global heat redistribution, described by a parameter $f$, which is related to the atmospheric circulation and varies between $f=0$ (no heat distribution) and $f=1$ (full distribution, leading to an isothermal surface). While rocky planets with no or little atmosphere, like Mercury, have an extremely high day–night temperature contrast, planets with a thick, Venus-like atmosphere tend to be nearly isothermal. Intermediate cases, with up to $10$ bar atmospheres, conserve significant surface temperature gradients (e.g., Selsis et al. 2011). For a locked planet the highest and lowest temperatures occur at the substellar point (denoted in this work by $\theta=0$) and at its antipode ($\theta=180$), respectively. These temperatures can be written as (W18)
\begin{equation}
T_{min} = 278\,(Hf)^{1/4} {\rm \,K,  \, \,and}
\end{equation}
\begin{equation}
T_{max} = 394\,H^{1/4} (1- \frac{3}{4} f) ^{1/4} {\rm \,K.}
\end{equation}

The surface temperature distribution with a uniform global heat redistribution is given by (W18; Fig. 1 therein)
\begin{equation}
T( \theta )=\,415\,  [ H\, F(\theta ) ] ^{ 1/4} \,{\rm K},
\end{equation}
where
\begin{equation}
F(\theta ) = 
\begin{cases}
f/4+(1-f)\cos (\theta ) &  \,\,0\leq\theta \leq90 \cr
f/4 & 90\leq \theta \leq 180.
\end{cases}
\end{equation}
Including local heat transport (e.g., by advection) leads to a differential equation, similar to the heat equation (Eq. 5 in W18) with a smoothed temperature distribution (see Fig. 1). If the advective heating is small, compared to the radiative heating due to the host star (e.g., on Earth at sea level advective heat transport is less than 1\% of the solar irradiation), the values of the substellar and antipode temperatures are approximately still given by Eqs. 2 and 3, respectively.

\begin{figure}[]
\includegraphics[width=1\linewidth]{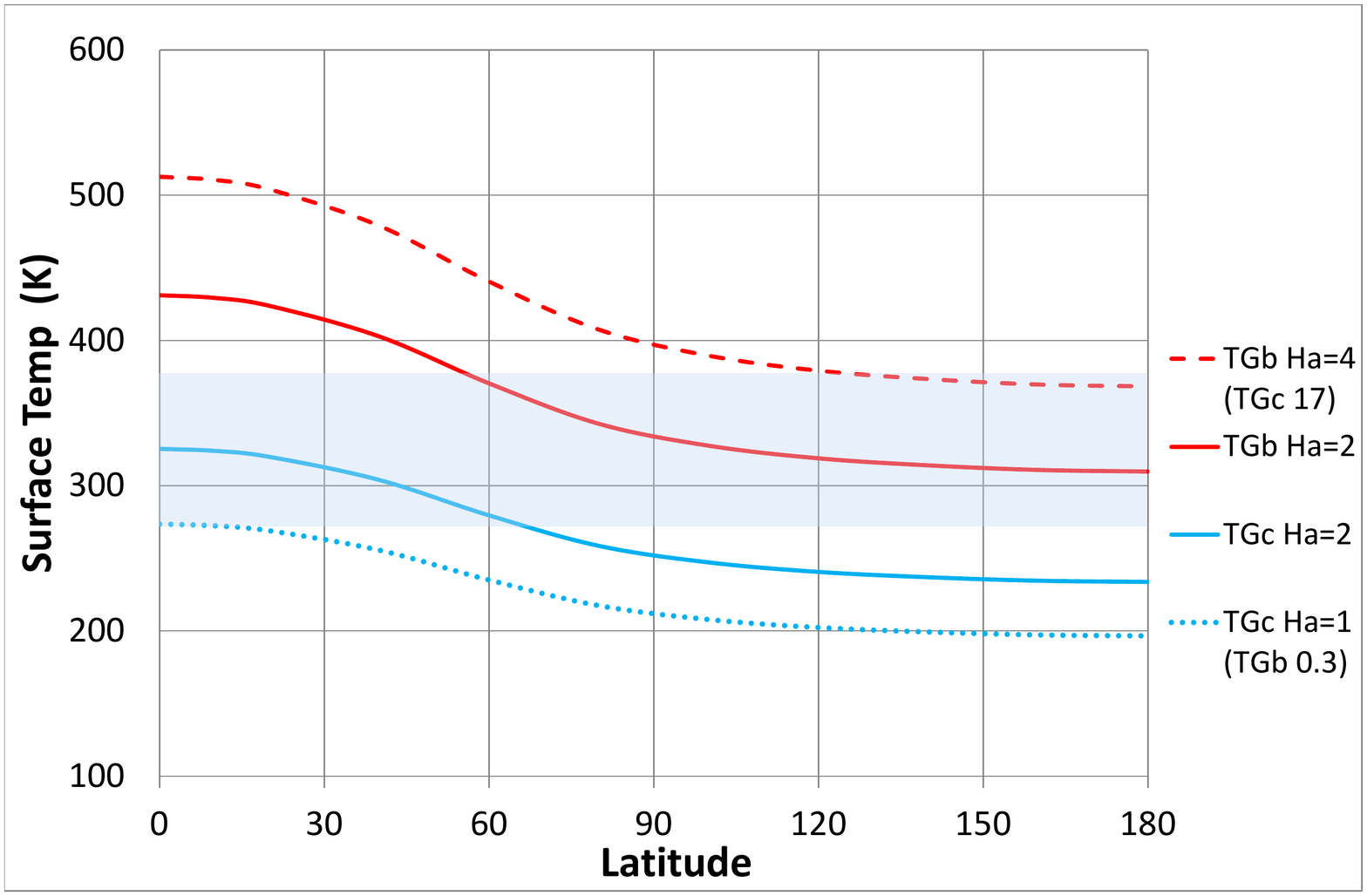}
\caption{Surface temperature profiles for TG's two planets. The global heat transport parameter is taken as $f=0.5$. The solid curves denote an atmospheric heating factor of $H_{atm}=2$. The lower dotted curve marks TGc with $H_{atm}=1$, and the upper dashed curve marks TGb with $H_{\rm atm}=4$. 
The upper dashed curve also corresponds to TGc with $H_{\rm atm}=17$, and the lower dotted curve to TGb with $H_{\rm atm}=0.32$
The temperature range of liquid water at $1$ bar is indicated by the shaded blue area.\label{fig:1}}
\end{figure}

Figure 1 shows the surface temperature distribution for the two planets TGb and TGc, for several values of the atmospheric heating as indicated on the legend on the right. In the range $0.3<H_{atm}<17$ there is a habitable region on some part of the surface of one or both planets. At $H_{atm}=1$ TGc has a narrow habitable region at the substellar (permanent day) point, at $H_{atm}=4$ TGb has a narrow habitable region at the substellar antipode (permanent night), and at $H_{\rm atm}=2$ both planets have a wide range of habitable latitudes, TGb between $60$--$180$ and TG $0$--$80$. The upper and lower curves also represent the two extreme ends of the habitability range of the atmospheric heating in the Teegarden's Star system. The upper dashed curve is also the temperature distribution of TGc with $H_{\rm atm}=17$, while the lower dotted curve also represents TGb with $H_{\rm atm}=0.32$. These numbers agree with the boundaries indicated in Table 1 and Fig. 2, for $f$=0.5. 

\subsection{The habitability range of the atmospheric heating}

The range of temperatures allowing liquid water on at least part of the planet surface could vary between freezing and the minimal moist greenhouse temperature ($\sim340$\,K) or higher, according to the atmospheric pressure and composition. The results do not strongly depend on this choice (W18), so we take this range as $273K<T<373$\,K. This temperature range defines the "habitability range" of the heating parameter. It extends between the lowest value, for which the substellar temperature is $273$\,K, and the highest value for which the substellar antipode is at $373$\,K (or $\sim340$\,K for a more conservative range). This gives (W18)
\begin{equation}
0.23\,(1 - \frac{3}{4} f)^{-1} < H < 3.2\,f^{-1} .
\end{equation}
Equations 1 and 6 give a relation between the planet's distance from the host star ($a$, in units of astronomical unit) and the atmospheric heating factor. For TG’s luminosity ($7.3\, 10^{-4}$\,L$_{\odot}$, Schweitzer et al. 2019) and $f=0.5$ we get
\begin{equation}
500\,a^2 < H_{atm} < 8770\,a^2 .
\end{equation}
Equation 6 may also be written in terms of the insolation and the atmospheric heating,
\begin{equation}
0.23\,(1 - \frac{3}{4} f)^{-1}\,s^{-1} < H_{atm} < 3.2\,f^{-1}\,s^{-1} .
\end{equation}

\begin{figure}[]
\includegraphics[width=1\linewidth]{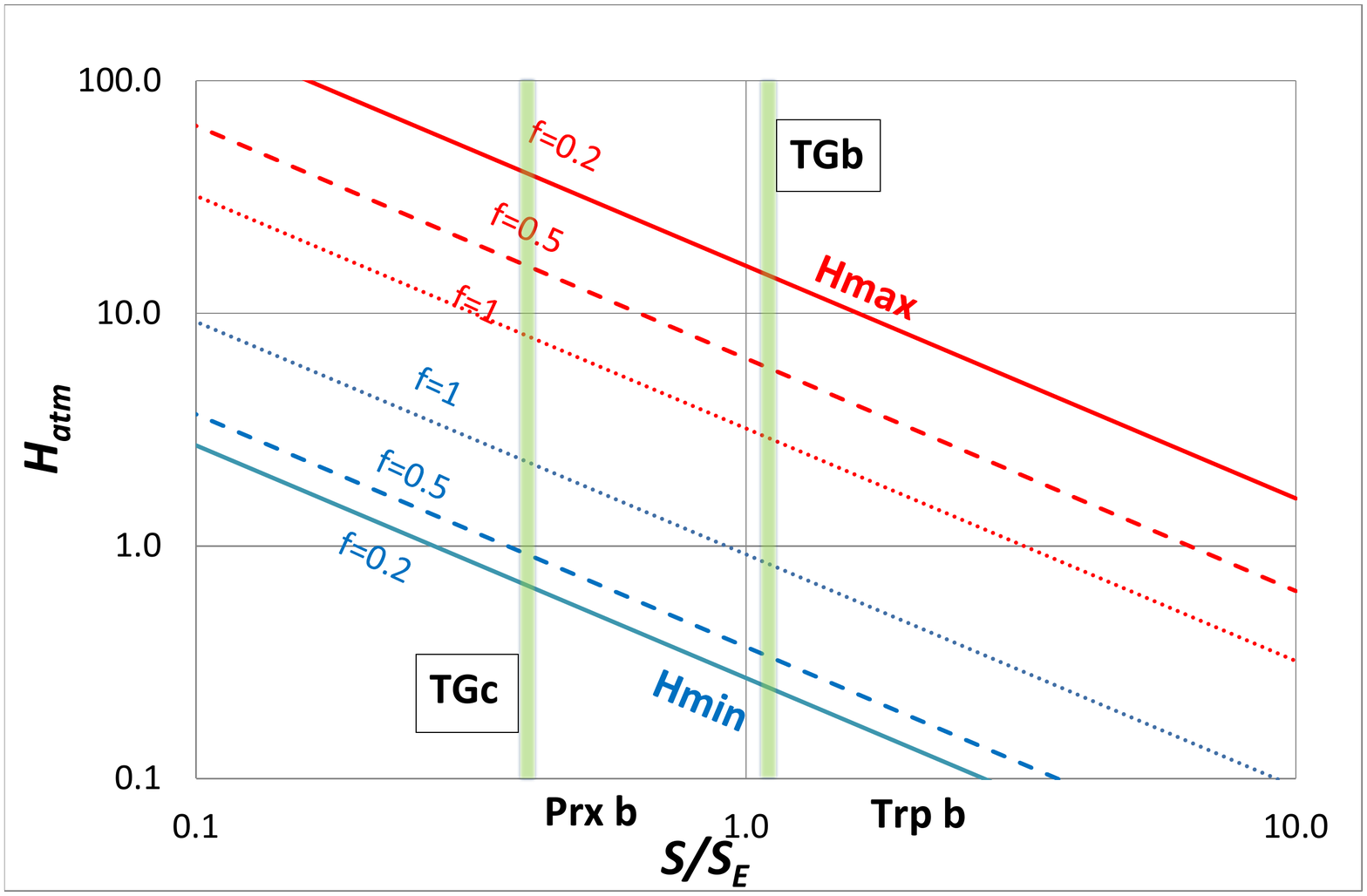}
\caption{Upper (red) and lower (blue) habitable boundaries of the atmospheric heating factor vs. insolation, for three values of the heat redistribution parameter: $f=0.2$ (solid), $0.5$ (dashed), and $1$ (dotted). The locations of TGb and TGc are marked with vertical green shaded strips. The locations of Proxima b and Trappist-1 b are marked on the x-axis. \label{fig:2}}
\end{figure}

The corresponding ranges of the atmospheric heating factor can be seen in Fig. 2, which shows the habitability boundaries of the atmospheric heating vs. the insolation (calculated with TG's luminosity). The biohabitable ranges for the planets are marked by vertical stripes. For a given insolation, the range corresponding to Eq. 6 is the vertical span between the lines of the highest habitable atmospheric heating factor (red curves) and the minimal value (blue) for a given value of the redistribution parameter $f$. Higher $f$ values (more atmospheric circulation causing heat distribution) narrow the range of habitability, as they either make the substellar point of planets at the HZ outskirts too cold, or make the substellar antipode of planets inward of the "traditional" HZ too hot.

The stellar flux on the two planets is $S({\rm TGb})\sim1.15$\,S$_{\oplus}$ and $S({\rm TGb})\sim0.37$\,S$_{\oplus}$, with an uncertainty of $\sim10$\,\%. This uncertainty comes mainly from the large systematic uncertainty (of $\sim10$\,\%) on $T_{\rm eff}$ of the star (Z19). Table 1 shows the habitability ranges in $H_{\rm atm}$ for each of the planets for two choices of heat circulation: $f=1$ (isothermal surface) and $f=0.5$, and for three cases of greenhouse upper limit: conservative, $T_{max}\sim340$\,K, corresponding to maximal moist greenhouse runaway (Kasting et al. 1993), optimistic, $T_{max}\sim373$\,K, and the highest bio-habitability limit (W18), $T_{max}\sim400$\,K. For $T_{max}\sim340$\,K the upper limits on $H_{\rm atm}$ will be smaller than those for $T_{max}\sim373$\,K by a factor of $\sim (340/373)^4 = 0.69$. Similarly, $T_{max}\sim 400$\,K, applicable to a higher atmospheric pressure, gives the upper $H_{\rm atm}$ limits larger by a factor of $\sim1.32$.

\begin{table*}[]
\centering
\caption{Habitable $H_{atm}$ ranges for each planet, for different choices of $f$ and $T_{max}$.} \label{tab:1}
\begin{tabular}{ccccc}
\tablewidth{0pt}
\hline
\hline
$T_{max}$ & $f=0.5$, TGb & $f=0.5$, TGc & $f=1.0$, TGb & $f=1.0$, TGc\\
\hline
$340$\,K & $0.32-3.7$ & $1-12$ & $0.8-1.9$ & $2.5-5.8$ \\
$373$\,K & $0.32-5.6$ & $1-17$ & $0.8-2.8$ & $2.5-8.6$ \\
$400$\,K & $0.32-8.0$ & $1-25$ & $0.8-4.0$ & $2.5-12$ \\
\hline
\end{tabular}
\end{table*}

Even in the conservative case, with the narrowest range ($f=1$), TGb can be habitable for an atmospheric heating of $0.7$--$1.6$ that of Earth, while for TGc the range is $2.2$--$5.0$. In the optimistic case and $50$\,\% heat circulation, the habitability range is significantly wider, $0.3$--$4.8$ for TGb and $0.9$--$15$ that of Earth, for TGc. Note that with the definition of $H_{atm}$ in W18, Earth's atmospheric heating factor is $H_{atm}=1.15$, so to compare with Earth the numbers in Table 1 need to be divided by $1.15$. For a lower heat circulation these limits are even wider, as can be seen in Fig. 2. We conclude that within the model and moderate circulation, at least one of the TG planets may be habitable for an atmospheric heating in the range $0.3$--$15$, that of Earth. 

%% =====================================================
\section{Discussion} \label{sec:3}

The habitability ranges we have drawn above and the redistribution parameter {\it f} assume the planets to have an atmosphere of some kind (unless {\it f=0}). However, tidally locked planets in the HZ of late M-dwarfs are known to have additional challenges on their way to become truly habitable planets. Here we discuss these items. 

\subsection{Early Atmospheric Erosion by Wind, Flares, and XUV flux}

It has been argued that the high levels of X-ray and ultraviolet (XUV) radiation and stellar winds may cause atmosphere erosion of small HZ planets of late M dwarfs (Lammer et al. 2007; Heller, Leconte and Barnes 2011; Leconte, et al. 2015; Lingam and Loeb 2017; Tilley et al. 2019). For a close-in planet, the planetary magnetic moment is strongly reduced by tidal locking. Hence, the planet is not protected by an extended magnetosphere against stellar winds, which at the small orbital distance of the planet is much denser than at larger orbital distances. Also the XUV stellar flux is contributing to atmospheric erosion. The current XUV fluxes and probably also wind of Teegarden's Star are relatively low, but they were probably higher in the past, when the star was more active for several Gyr. Recently, Fleming et al. (2019) modeled the long-term XUV luminosity of TRAPPIST-1 (which has a similar mass and age to TG), finding that TRAPPIST-1 has maintained high activity and $L_{\rm XUV}/L_{\rm bol}\approx0.001$ for several Gyr. In addition, erosion is much higher during stellar flares, which were more frequent in the younger TG. However, detailed numerical simulations including extreme ultraviolet fluxes and stellar winds have been carried out for Proxima b and the TRAPPIST-1 planets, which have somewhat similar hosts and insolations to the TG's planets. These simulations have shown that an "Earth-like" 1 bar atmosphere might be retained over Gyr timescales in certain circumstances (Dong et al. 2017; 2018). 

Atmospheric erosion by stellar wind may be inhibited by a planetary magnetic field. Although the magnetic field of old, locked planets like TGb and TGc may be weak, it has probably been stronger during earlier epochs, when the host star was more active and had stronger winds and outbursts. The tidal locking and hence loss of magnetosphere depends on the planetary composition and is of the order of a few Gyr (Griessmeier et al. 2009). Furthermore, recent numerical and analytic calculations show that even without magnetic fields planets can have fairly low atmospheric escape rates and fluxes of ionizing radiation at the surface (e.g., Lingam 2019).

Also the evolution of the host, leading to a calming down of the stellar activity is of the same order, depending on the rate of magnetic braking of the star, leading to the stellar spin decay, and hence to lower activity levels. Furthermore, the eroded atmosphere may be compensated by accretion of a secondary atmosphere or outgassing during the later, calmer phases in the host’s evolution, which also last much longer. Additionally, a massive prime atmosphere could survive the extended erosion during the energetic early evolution of M-dwarfs (e.g., Tian 2009). As TGb and TGc are more massive than Earth, by up to a factor of $\sim2$, they could potentially have thicker initial atmospheres and possibly also stronger magnetic fields to protect them.

\subsection{Atmospheric Collapse and Water Trapping on Locked Planets}

Locked or nearly locked planets have been shown to have peculiar climates, which may affect their habitability (e.g., Kopparapu et al. 2016; Checlair et al. 2017). Simulations show that locked or synchronous habitable-zone planets of M-type stars may support liquid water oceans (e.g., Del Genio et al. 2019). This depends not only on the irradiation from the host star, but to a large extent on the planet's atmosphere. Global circulation models (GCMs) using radiative transfer, turbulence, convection, and volatile phase changes can be used to calculate the conditions on planets, given the properties of their atmospheres. Such 3D climate models of M-dwarf planets suggest the presence of liquid water for a variety of atmospheric conditions (e.g., Pierrehumbert 2011; Wordsworth 2015). Climate modeling studies have shown that an atmosphere only $10$\,\% of the mass of Earth's atmosphere can transport heat from the day side to the night side of tidally locked planets, enough to prevent atmospheric collapse by condensation (Joshi et al. 1997; Scalo et al. 2007; Tarter et al. 2007; Heng and Kopparla 2012). Also Wordsworth (2015) has demonstrated that atmospheric heat redistribution of that order is enough on tidally locked rocky planets.

On locked planets the water may be trapped on the night side (e.g., Leconte 2013), but on planets with enough water or geothermal heat, part of the water remains liquid (Yang et al. 2014). Geothermal heating is assumed to be negligible compared to radiative heating, which is likely for TG’s planets given their age. 3D GCM simulations of planets in the habitable zone of M-dwarfs support scenarios with surface water and moderate temperatures (Yang et al. 2013; 2014; Leconte et al. 2015; Owen and Mohanty 2016; Turbet et al. 2016; Kopparapu et al. 2016; Wolf 2017).

\subsection{Water Loss at Early Evolutionary Stages}

In the highly luminous pre-main-sequence phase of M-dwarfs, which lasts about $10^9$ yr for late-type M-dwarfs like Teegarden's Star, the planets may be susceptible to losing a large part of their original water content because of the enhanced stellar activity. During this early runaway phase, photolysis of water could result in hydrogen/oxygen escape to space of large quantities of water, up to several times that of all the Earth's oceans (Tian et al. 2014; Luger and Barnes 2015). However, this early on water loss does not preclude the existence of surface water during the later, calmer stages (e.g., Gale and Wandel 2017; Wandel \& Gale 2019). Being well within the frost line, most of the water of both planets was probably acquired after their formation via bombardment by comets and asteroids, and could probably be reacquired by late bombardment.

%% =====================================================
\section{Summary} \label{sec:4}

The planet candidates of Teegarden's Star, TGb and TGc, are likely to support liquid water on at least part of their surface for a wide range of possible atmospheres, characterized by their atmospheric heating factor (product of the greenhouse effect, screening and albedo) and global heat circulation. At least one of the TG planets may be habitable for an atmospheric heating in the range 0.3-15, that of Earth. As demonstrated by detailed  numerical calculations of similar planets, Teegarden's Star present calmness and old age favor the retaining or reproduction of a sufficiently massive atmosphere, with a heating factor within the habitability range.

%% =====================================================

\acknowledgments

We thank the anonymous referee for their swift response and the very useful comments and suggestions. L.T.-O. acknowledges support from the Israel Science Foundation (grant No. 848/16).

%% =====================================================

%% This command is needed to show the entire author + affiliation list when


\end{document}